\documentclass[sigconf, nonacm, x11names]{acmart}

\usepackage[english]{babel}
\usepackage{blindtext}
\usepackage{changepage}
\usepackage{graphicx}
\graphicspath{ {Figures/} }
\usepackage{caption}
\usepackage{subcaption}
\usepackage{stackengine}
\usepackage{float}

\usepackage{amssymb}
\usepackage{url}
\usepackage{listings}
\usepackage{enumitem}

\definecolor{todocolor}{rgb}{0.8,0,0}
\definecolor{darkgreen}{rgb}{0.09, 0.45, 0.27}
\definecolor{neilcolor}{rgb}{0.7, 0.2, 0.27}

\definecolor{donecolor}{rgb}{0,0,0.8}

\newcommand{\HIDE}[1]{}

\setlist[description]{style=nextline}

\begin{document}

\title{Cryptographic Data Exchange for Nuclear Warheads}


\author{Neil Perry}
\authornote{Both authors contributed equally to the paper.}
\affiliation{%
    \institution{Stanford University}
    \country{}
}
\author{Daniil Zhukov}
\authornotemark[1]
\authornote{The author was employed by Stanford University while conducting the work described in this paper.}
\affiliation{%
    \institution{University of California, Berkeley}
    \country{}
}

\begin{abstract}
Nuclear arms control treaties have historically focused on strategic nuclear delivery systems, indirectly restricting strategic nuclear warhead numbers and leaving nonstrategic nuclear warheads (NSNWs) outside formal verification frameworks. This paper presents a cryptographic protocol for secure and verifiable warhead tracking, addressing challenges in nuclear warhead verification without requiring intrusive physical inspections. Our system leverages commitment schemes and zero-knowledge succinct non-interactive arguments of knowledge (zkSNARKs) to ensure compliance with treaty constraints while preserving the confidentiality of sensitive nuclear warhead data. We propose a cryptographic “Warhead Passport” tracking system that chains commitments to individual warheads over their life cycle, enabling periodic challenges and real-time verification of treaty compliance. Our implementation follows real-world treaty constraints, integrates U.S. and Russian dual-hash combiners (SHA-family \& GOST R 34.11 family) for cryptographic robustness and political constraints, and ensures forward security by preventing retroactive data manipulation. This work builds on policy research from prior arms control studies and provides a practical foundation for implementing secure, auditable NSNW verification mechanisms.
\end{abstract}

\maketitle

\section{Introduction}

Nuclear arms control treaties have been a cornerstone of global security for over half a century, reducing nuclear arsenals and providing mechanisms to verify those limits and reductions. However, these treaties have historically focused on strategic nuclear delivery systems, indirectly restricting strategic nuclear warhead numbers and leaving nonstrategic nuclear warheads (NSNWs) outside formal verification frameworks. Future advancements toward nuclear arms reductions and disarmament will require addressing these weapons within arms control treaties, which is challenging because NSNWs are more difficult to define and track, are highly mobile, and are susceptible to modernization and covert redeployment. Addressing these challenges requires a robust verification system that ensures compliance while maintaining the confidentiality of sensitive military data.

In this work, we present a cryptographic protocol for secure and verifiable nuclear warhead tracking, enabling real-time treaty compliance monitoring without intrusive physical inspections. Our approach introduces a "Warhead Passport" system that chains together cryptographic commitments over the life cycle of each warhead. This system allows for periodic data challenges and real-time verification of compliance through zero-knowledge succinct non-interactive arguments of knowledge (zkSNARKs), ensuring both data integrity, secrecy, and treaty compliance.

A central component of our system is the data challenge process, which is critical for building confidence and trust between nations. The ability to challenge past commitments and verify selective information from selectively revealed inventory information ensures that each party can confirm the other’s adherence to treaty constraints and build their confidence in their knowledge of the other’s inventory. This ongoing verification mechanism fosters transparency while preserving security, providing a scalable and enforceable means of confirming compliance. Trust in the integrity of these inventory-sharing statements is essential for progress in disarmament efforts, as confidence in accurate and truthful reporting is a necessary step toward reducing and ultimately eliminating nuclear weapons. This describes the technical system behind the policy work done in Pomper~et al.~\cite{nonproliferationOP55Everything}.

The key contributions of our work are as follows:
\begin{description}
    \item[Formalization of Treaty Rules] We translate real-world treaty constraints into a structured cryptographic framework, incorporating explicit rules for warhead movement, status updates, and personnel verification.
    \item[Warhead Passport System] We introduce a novel method for tracking warheads using chained commitments that enforce forward security, preventing retroactive data manipulation.
    \item[zkSNARK-Based Compliance Verification] By leveraging zkSNARKs, we provide a mechanism for verifying treaty adherence without revealing classified warhead details, addressing the confidentiality concerns that have historically hindered warhead-level verification.
    \item[Data Challenge Process for Trust and Confidence Building] Our system enables periodic data challenges, allowing nations to selectively reveal and verify past commitments. This is critical for ensuring trust in inventory-sharing statements, verifying treaty compliance, and advancing nuclear disarmament efforts.
    \item[Cryptographic Robustness via Dual-Hash Combiners] Our system integrates both SHA-family and GOST R 34.11 family hash functions, ensuring resilience against single-standard vulnerabilities and addressing political concerns regarding cryptographic trust.
    \item[Standards-Conforming Implementation] We provide a fully open-source implementation of our system, written in Python and ZoKrates. Additionally, we provide a complete implementation of the GOST R 34.11.94 hash function standard within the ZoKrates framework, making zkSNARK-based verification more accessible for projects needing to conform to Russian standards. These enable follow-up research in cryptographic arms control mechanisms. The code can be found at \url{https://github.com/NeilAPerry/Warhead-Tracking-System}.
\end{description}

By applying these techniques, we demonstrate that the solution of nuclear warhead verification can be streamlined by applying treaty rules to sequential database updates. This structured approach allows for efficient warhead data exchange verification while maintaining strict secrecy, making it feasible for real-world deployment in arms control agreements. Our protocol provides a scalable and enforceable foundation for future NSNW verification mechanisms, bridging the gap between cryptographic theory and international security policy while reinforcing trust—a key component in the movement toward nuclear disarmament.
\section{Background}

\subsection{History of Nuclear Verification \& Challenges}
\label{sec:nuclear_history}

Nuclear arms control treaties between the United States and the Soviet Union/Russia have largely focused on limiting or reducing strategic nuclear delivery systems. While suitable for monitoring relatively large objects like missiles and bombers, verification regimes established by past treaties are insufficient for monitoring nuclear warheads directly, which will be required by any future agreement that seeks to reduce entire nuclear warhead stockpiles.

Strategic nuclear weapons are considered to be weapons that can be delivered to distances over 5,500 kilometers and thus include: intercontinental ballistic missiles (ICBMs), submarine-launched ballistic missiles (SLBMs), and heavy bombers. The U.S.-Russian New Strategic Arms Reduction Treaty (\textit{New START}) caps both countries' strategic nuclear arsenals at 1,550 deployed warheads, 700 deployed launchers for those weapons, and 800 deployed and non-deployed launchers \cite{newstart2010}. While the treaty officially restricts the number of deployed strategic nuclear warheads, it does so by attributing a specific number of nuclear warheads to a given delivery system rather than directly monitoring the warheads themselves. 

The treaty mandates that both countries provide \textit{advance notifications} related to movements of missiles and bombers and other relevant events. In order to verify the data exchanged through those notifications and monitor compliance with the overall treaty limits, New START and its predecessors have relied on \textit{physical inspections} of missiles and bombers, \textit{national technical means of verification} (NTM) that include satellite imagery, and \textit{intelligence gathering}.

In contrast, \textit{nonstrategic nuclear warheads (NSNW)} have largely remained outside formal verification frameworks \cite{woolf2006nonstrategic}. Non-strategic nuclear weapons can be delivered at shorter ranges and  require smaller delivery systems than ICBMs, SLBMs, or heavy bombers. The United States government estimates that Russia possesses between 1,000 and 2,000 non-strategic nuclear warheads and a wide range of dual-capable delivery systems to launch those warheads\cite{RussianNSNW2024}. Washington has long sought to negotiate an arms control treaty that reduces that number. However, a treaty or an agreement addressing NSNW runs into challenges of monitoring compliance; the lack of direct inspections for warheads themselves creates concerns over \textit{frequent and undetected stockpile changes, covert deployments, and possible treaty violations}.

James Fuller~\cite{fullerverification} explains how conventional verification (inspections, NTM like satellite imagery, and intelligence gathering) works, but why it falls short for warhead-level accounting. Traditional verification methods suffer from \textit{several limitations}:
\begin{itemize}
\item \textbf{Physical Inspections}: Require mutual trust and access to sensitive military facilities, which nations may refuse.
\item \textbf{Satellite Surveillance}: Can detect movement but cannot distinguish between nuclear and conventional assets.
\item \textbf{Human Intelligence (HUMINT)}: Prone to deception and misinterpretation.
\end{itemize}

The U.S. Government Accountability Office \cite{verifying_challenges} further underlines the practical challenges (size, mobility, dual-use ambiguity) that motivate a secure data exchange approach beyond traditional methods. Since the advent of nuclear weapons in 1945, arms control agreements have attempted to reduce risk through transparency, verification, and limitations on deployment. However, many of these treaties have been shaped as much by the verification tools available at the time as by the political will to act. Below, we trace the key milestones in arms control history and highlight challenges that remain unresolved or imperfectly addressed, laying the groundwork for why cryptographic tools are now poised to make a difference.

\begin{description}
    \item[1945 – Nuclear weapons first used] The United States drops atomic bombs on Hiroshima and Nagasaki. This event demonstrates the devastating power of nuclear weapons, triggers the nuclear arms race, and creates a long-term need for mitigating the nuclear danger~\cite{swift2009soviet}.
    \item[1970 – Non-Proliferation Treaty (NPT) enters force] Aims to prevent the spread of nuclear weapons while promoting peaceful uses of nuclear energy. Signatories to the Treaty pledge to engage in negotiations to restrain the nuclear arms race and achieve nuclear disarmament under Article VI~\cite{npt1968}.
    \item[1972 - Strategic Arms Limitation Talks (SALT I) Interim Agreement and Anti-Ballistic Missile (ABM) Treaty signed] Places limits on U.S. and Soviet strategic offensive and defensive systems. These agreements are the outcome of the first U.S.-Soviet bilateral nuclear arms control negotiations and open the door for future treaties and agreements. SALT I introduces the concept of \textit{National Technical Means of Verification}, including satellite imagery, that both parties have the right to use for monitoring each other's compliance with the agreement~\cite{salt1abm}.
    \item[1987 – Intermediate-range Nuclear Forces (INF) Treaty signed] Eliminates intermediate-range missiles. Verification relies on on-site inspections and perimeter monitoring. Missiles are the primary treaty-banned item; there is no provision for monitoring warheads and verifying their identity or sensitive components~\cite{inf1987}.
    \item[1991 – Strategic Arms Reduction Treaty (START I) signed] Focuses on reducing strategic delivery systems and deployed warheads. Warheads themselves are not tracked individually, but rather counted via their association with strategic delivery systems~\cite{start1_1991}.
    \item[2010 – New START Treaty signed] Updates START I to further reduce deployed strategic arms to 1,550 deployed warheads each. New START updates START I counting rules, and its verification mechanisms are similar to past strategic arms control treaties, including telemetry and inspections. The treaty does not cover NSNW, and both sides fail to negotiate a new treaty addressing all nuclear weapons in the subsequent years~\cite{newstart2010}.
\end{description}

These milestones demonstrate a recurring theme: nuclear arms control treaties have thus far focused on delivery systems because those items are relatively easy to observe, enabling effective compliance monitoring regimes. A treaty or agreement addressing warheads directly would require more intrusive verification measures that run against the nuclear-weapon states' security and secrecy considerations. This circumstance presents the challenge that Jane Vaynman terms the ``transparency-security tradeoff'' \cite{vaynman2021}. The inability to verify warhead-level data without compromising national security has thus far prevented agreements that impose restrictions of greater scope on nuclear arsenals.

\subsection{Cryptographic Advances Enabling Private Verification}

While arms control treaties have historically evolved alongside geopolitical needs, their verification mechanisms have often lagged due to the limits of available technology. From the INF to the New START, verification relied on physical inspections, technical means, and intelligence—but these methods were often too blunt or intrusive to track warheads directly, especially nonstrategic nuclear warheads. Modern cryptographic tools now make it possible to enforce treaty rules while preserving secrecy, directly addressing the trust-versus-transparency tradeoff that has historically stalled progress. 

\begin{figure*}[h]
    \centering
    \includegraphics[width=1\linewidth]{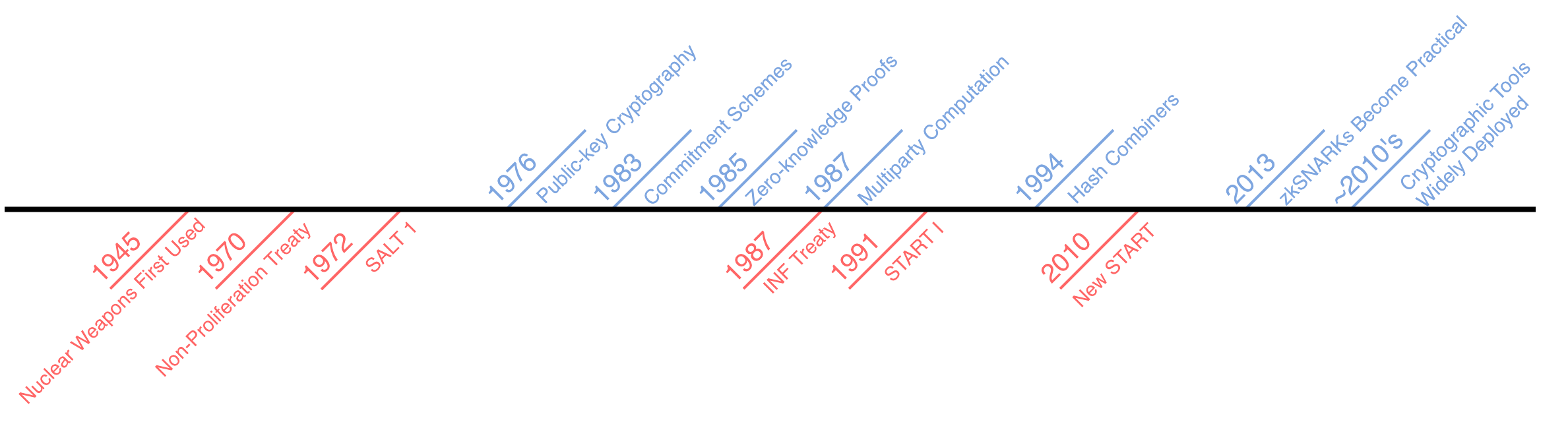}
    \caption{Timeline showing key events in arms control and cryptography.}
    \label{fig:timeline}
\end{figure*}

The timeline below highlights key cryptographic breakthroughs and how they now align with long-standing arms control needs, as detailed in Section~\ref{sec:nuclear_history} (also see Figure~\ref{fig:timeline}). Each step in cryptographic development brings us closer to the kind of fine-grained, real-time, and privacy-preserving verification envisioned in our “Warhead Passport” system.

\begin{description}
    \item[1976 – Public-key cryptography introduced] Diffie-Hellman and RSA establish the foundations for secure digital communication. These tools enable digital signatures, authenticated data sharing, and many other necessities in the warhead tracking system~\cite{diffie2022new}.

    \item[1980s – Commitment schemes formalized] Cryptographers develop methods for committing to data without revealing it, only later allowing selective disclosure. This directly supports the notion of sealed warhead declarations, revealed only when challenged~\cite{blum1983coin,micali1987play}.

    \item[1985 – Zero-knowledge proofs (ZKPs)] Introduced by Goldwasser, Micali, and Rackoff, ZKPs allow one party to prove that data satisfies a rule (e.g., that a warhead update follows treaty constraints) without revealing the underlying data itself~\cite{goldwasser2019knowledge}.

    \item[1980s – Secure multiparty computation (MPC)] MPC allows multiple distrustful parties to jointly compute functions over their private inputs without revealing them. Though initially inefficient, this laid the theoretical groundwork for collaborative but confidential treaty verification~\cite{yao_protocols,micali1987play,completeness_theorems}.

    \item[1994 – Hash combiners introduced] Address the challenge of geopolitical distrust in cryptographic standards. For example, Russia may prefer a hash function from GOST R 34.11, while the U.S. uses SHA-256. Hash combiners allow both to be used in parallel, ensuring robustness as long as one remains secure~\cite{preneel1993analysis,herzberg2005tolerant,boneh2006impossibility}.

    \item[2013 – zkSNARKs become practical] Parno~et~al.~\cite{pinocchio_2013} introduce the first efficient zero-knowledge Succinct Non-interactive Argument of Knowledge, enabling compact, constant-size proofs for complex computations. This makes real-time verification of treaty updates feasible.

    \item[2010s–2020s – Advanced cryptographic tools reach real-world scale] zkSNARKs and efficient MPC are deployed in systems like Zcash~\cite{sasson2014zerocash}, demonstrating that these techniques can operate at national or even global scale. Their success in large, high-stakes financial applications highlights their readiness for adoption in arms control.
\end{description}

These developments enable a new class of verification systems. Our work applies these tools in the context of nuclear arms control, where each warhead’s state evolves as a series of updates. Monitoring compliance involves checking that each update complies with treaty rules, much like enforcing constraints on row transitions in a database. This structure maps naturally onto cryptographic primitives: commitments hide data, zkSNARKs prove compliance via conforming to rules and transitions between database entries, and challenge protocols verify specific facts over time without leaking unnecessary information. Together, these advances enable a warhead-level tracking system that was previously impossible.

\subsection{DIAMONDS and AICMS as Trusted Data Sources}

While cryptographic protocols can enforce treaty compliance without revealing sensitive information, any verification system ultimately relies on accurate underlying data. In the context of nuclear arms control, the U.S. and Russia already operate independent inventory management systems: the Defense Integration and Management of Nuclear Data Services (\textit{DIAMONDS}) in the United States and the Automated Inventory Control and Management System (\textit{AICMS}) in Russia~\cite{moon_2022}. These national-level databases track warhead movements, maintenance, and decommissioning activities, and they serve as the authoritative sources for internal decision-making.

Although not originally designed for bilateral verification, DIAMONDS and AICMS are kept accurate due to the high cost of internal errors and the consequences of maintaining fake secondary databases:
\begin{enumerate}
\item \textbf{Internal Security Risks}: Misalignment between real-world warhead locations and internal databases increases the risk of misplaced or unaccounted-for weapons, which could lead to theft, unauthorized use, or terrorism.
\item \textbf{External Surveillance}: Adversaries and allies alike employ independent intelligence tools—such as satellites and signal intercepts—that could expose inconsistencies, damaging diplomatic credibility.
\end{enumerate}

Because of these incentives, both parties have strong reason to maintain accurate records, making these systems well-suited as the private ground truths for cryptographic verification. Our approach builds on this foundation by treating each inventory system as a private data source that emits cryptographic commitments when an event occurs—such as the movement, inspection, or dismantlement of a warhead. These event-driven updates naturally lend themselves to zero-knowledge proofs, which verify that each state transition (e.g., before and after an event) complies with rules derived from treaty language.

Each event is modeled as appending an update row to a ledger. Rather than revealing the content of the change, our system generates a zkSNARK proving that the transition from the previous state to the new state is valid under agreed-upon constraints—such as ensuring that a warhead is not moved to an undisclosed or illegal location. This structure enforces treaty rules over time without ever disclosing classified operational data.

In this way, DIAMONDS and AICMS provide high-integrity data sources, and our cryptographic system transforms them into verifiable but private instruments of international trust.

\subsection{Contributions}
Existing work \textit{does not fully address the specific requirements of nuclear verification}. Specifically, most open-source research focuses on ways to streamline direct warhead inspections and verify warhead identity. Those measures are important components of a future agreement limiting or reducing warhead numbers directly, but the question of how countries can exchange relevant information about their warhead stockpiles without betraying sensitive details and technical characteristics is not fully solved. 

Notably, previous work addressed the issue of sharing full warhead stockpile declarations with the help of modern cryptography. A 2005 National Academies of Science report on nuclear verification highlights cryptographic techniques like hashing as useful means of exchanging "escrows" of full stockpile declarations between treaty sides that can be gradually revealed over time to build confidence in the entire commitment \cite{NAS2005}.

Our warhead tracking system builds on this approach and develops novel procedures for exchanging granular data on nuclear warhead stockpiles by:
\begin{itemize}
\item \textbf{Chaining commitments together for individual warhead tracking.}
\item \textbf{Using dual-hash combiners (SHA-family \& GOST R 34.11-family) to eliminate dependency on a single trusted cryptographic standard.}
\item \textbf{Integrating zkSNARK proofs to ensure all commitments comply with treaty constraints.}
\end{itemize}

This combination allows \textit{frequent updates (several per day)} while maintaining a high level of secrecy and verifiability, ensuring compliance without revealing classified information. Further, Pomper et al. \cite{op64_pomper_2024} highlights the U.S. government's long-standing goal of directly limiting warheads and demonstrates the system's utility in achieving this goal, showing that the policy community values the use case of such a system.

\subsection{Development Methodology}
The warhead tracking system outlined above was proposed in a 2021 James Martin Center for Nonproliferation Studies (CNS) report by Pomper et al. \cite{nonproliferationOP55Everything}. While developing the system, the CNS team (including the authors of this paper) engaged in extensive consultations with U.S. policymakers, technical and policy-oriented subject-matter experts in nuclear weapons, and leading cryptographers.

Specifically, the report's authors consulted with the following (non-exhaustive) list of institutions relevant to the U.S. nuclear weapons policy and research:
\begin{itemize}
    \item \textbf{White House Office of Science and Technology Policy}
    \item \textbf{State Department}
    \item \textbf{Defense Department}
    \item \textbf{National Nuclear Security Administration}
    \item \textbf{Department of Energy national laboratories}
    \item \textbf{National Academies of Sciences, Engineering, and Medicine}
\end{itemize}

These consultations and discussions with U.S. Allies, partners, and non-governmental experts shaped the development of the warhead tracking system. The CNS demonstrated a proof-of-concept prototype of the system to the U.S. State Department in 2023 \cite{brownStateDemo}. A comprehensive overview of the system's technical characteristics follows below.
\section{Preliminaries}

This section introduces the formal notation and cryptographic primitives used in our system: pseudorandom functions to generate secure randomness deterministically, hash combiners to ensure security when parties distrust each other’s cryptographic standards, commitment schemes to hide data until it can be revealed, and zero-knowledge proofs to enable private verification. We provide formal definitions and illustrative constructions where needed in order to understand their role in enabling secure treaty compliance.

\subsection{Pseudorandom Functions (PRFs)}

A function family \(\mathsf{PRF} = (\mathsf{PRFGen}, \mathsf{PRFEval})\) is a pseudorandom function \cite{goldreich1986construct} if:
\begin{itemize}
  \item \(\mathsf{PRFGen}\) is a probabilistic polynomial-time algorithm that takes as input a security parameter \(\lambda \in \mathbb{N}\) and outputs a key \(k \in \{0,1\}^\lambda\).
  \item \(\mathsf{PRFEval}\) is a deterministic polynomial-time algorithm that takes as input a key \(k \in \{0,1\}^\lambda\) and a domain element \(x \in \mathcal{X}_\lambda\), and outputs \(y \in \mathcal{Y}_\lambda\).
\end{itemize}

Throughout, we use the notation $PRF_k(\cdot)$ to denote $\mathsf{PRFEval}(k, \cdot)$, the evaluation of the pseudorandom function keyed by $k$.  The function family is pseudorandom if for every probabilistic polynomial-time distinguisher \(D\), there exists a negligible function \(\nu(\lambda)\) such that:

\[
\left|
  \Pr_{k \gets \{0,1\}^\lambda}\bigl[ D^{\mathsf{PRF}_k(\cdot)}(1^\lambda) = 1 \bigr]
  \;-\;
  \Pr_{f \gets \mathcal{F}_\lambda}\bigl[ D^{f(\cdot)}(1^\lambda) = 1 \bigr]
\right|
\;\;\le\;\;
\nu(\lambda),
\]

where \(\mathcal{F}_\lambda\) is the set of all functions mapping \(\mathcal{X}_\lambda\) to \(\mathcal{Y}_\lambda\). In our setting, PRFs are used to deterministically derive secure randomness for commitments and other aspects of the system.

\subsection{Cryptographic Combiners}

A cryptographic combiner is a mechanism that allows two parties to use different cryptographic primitives while ensuring security as long as at least one primitive is secure \cite{herzberg2005tolerant,boneh2006impossibility}. For example:
\begin{itemize}
\item $H_{1}$: A hash function chosen by party 1.
\item $H_{2}$: A hash function chosen by party 2.
\end{itemize}
The combined hash function is defined as:
\[
H(x) = H_{1}(x) || H_{2}(x)
\]
where $||$ denotes concatenation. This ensures security as long as at least one of the hash functions remains collision-resistant. This is particularly useful when ensuring that the constructions remains trusted regardless of a party's trust in the other party's choice.

\subsection{Commitment Schemes}

A \emph{commitment scheme} allows one party (the committer) to commit to a value while keeping it hidden, with the ability to reveal it later. Formally, a commitment scheme consists of two algorithms:
\begin{itemize}
\item $\text{Commit}(m, r) \rightarrow C$: Computes a commitment $C$ to a message $m$ using randomness $r$.
\item $\text{Open}(C, m, r) \rightarrow \{\text{true}, \text{false}\}$: Verifies whether $C$ is a valid commitment to $m$.
\end{itemize}
A secure commitment scheme satisfies:
\begin{itemize}
\item \textbf{Hiding}: Given $C$, an adversary cannot determine $m$.
\item \textbf{Binding}: Once $C$ is published, the committer cannot reveal a different $m' \neq m$ such that $\text{Open}(C, m', r) \rightarrow \text{true}$.
\end{itemize}

\subsubsection{Merkle Tree--Based Commitment with Inclusion Proofs}

A concrete realization of a commitment scheme can be obtained using a Merkle tree \cite{merkle2019protocols} over components of the message. Suppose the message $m$ is split into $n$ elements:
\[
m = (m_1, m_2, \dots, m_n).
\]
Each element is hashed individually,
\[
h_i = H(m_i), \quad i = 1, \ldots, n,
\]
where $H$ is a cryptographic hash function. These $h_i$ serve as the leaves of a binary Merkle tree. The tree is constructed by iteratively hashing the concatenation of each pair of child nodes until a single \emph{root} hash $C$ is obtained. This root $C$ is published as the commitment to $m$.

\paragraph{Proving and Verifying Inclusion.}
In addition to standard commitment properties (hiding and binding), this Merkle‐based construction efficiently supports \emph{inclusion proofs} for any subset of leaves. Let $L$ be a subset of indices corresponding to leaves $\{\,x_i : i \in L\}$ in the Merkle tree (where each $x_i$ is $H(m_i)$). An inclusion proof is a compact set of sibling hashes $\pi$ sufficient for a verifier to recompute the Merkle root from those specific leaves. We define two interfaces:

\[
\text{ProveInclusion}(L,\{x_i\}_{i \in L}) \,\rightarrow\, \pi, 
\]
which returns an aggregated set of sibling hashes (the proof) for leaves in $L$, and
\[
\text{VerifyInclusion}(\pi, \{x_i\}_{i\in L}, C) \,\rightarrow\, \{\text{true}, \text{false}\},
\]
which uses $\pi$ to rebuild the path(s) to the root, producing \(\text{true}\) if (and only if) the reconstructed root $\hat{C}$ equals the published commitment $C$.

\paragraph{Example: Aggregated Inclusion Proof for Multiple Leaves.}

Consider a Merkle tree with the structure:
\[
\begin{array}{c}
\text{root} \\
x_0 \quad x_1 \\
x_2 \quad x_3 \quad x_4 \quad x_5 \\
x_6 \quad x_7 \quad x_8 \quad x_9 \quad x_{10} \quad x_{11} \quad x_{12} \quad x_{13}
\end{array}
\]
Suppose we wish to prove that leaves $x_6$ and $x_{12}$ are included. One possible aggregated proof is
\[
\pi = \{\;H(x_7),\; x_3,\; x_4,\; H(x_{13})\}.
\]
To verify, the verifier uses $\{x_6, x_{12}\}$ plus the sibling hashes in $\pi$ to reconstruct the necessary internal nodes and ultimately compute a candidate root $\hat{C}$. The function $\text{VerifyInclusion}(\pi, \{x_6, x_{12}\}, C)$ outputs \(\text{true}\) if $\hat{C} = C$, and \(\text{false}\) otherwise. If it is \(\text{true}\), the verifier is convinced that both $x_6$ and $x_{12}$ were included in the committed structure without learning all other leaves.

\subsection{Zero-Knowledge Succinct Non-Interactive Arguments of Knowledge (zkSNARKs)}

zkSNARKs are cryptographic proofs that enable a prover to convince a verifier that a computation is performed correctly, without revealing any additional information about the underlying data.

\begin{itemize}
    \item \textbf{Zero-Knowledge:} The proof discloses nothing beyond the fact that the underlying statement is true.
    \item \textbf{Succinctness:} Proofs are compact. This makes them easy to send and allows for rapid verification, even for complex computations.
    \item \textbf{Non-Interactivity:} Once the proof is generated, no further interaction is needed between the prover and verifier, which simplifies the protocol.
    \item \textbf{Soundness:} It is computationally infeasible for a dishonest prover to generate a valid proof for a false statement, ensuring the integrity of the verification process.
\end{itemize}

Formally, given a statement \( S \) and a corresponding witness \( w \), the prover computes a proof \( \pi \) using a function $Prove$. The common reference string (CRS) is a publicly available set of parameters generated during a trusted setup phase and used by both the prover and verifier.
\[
\pi \leftarrow \text{Prove}(S, w, \text{CRS})
\]
The verifier then checks the proof using:
\[
\text{Verify}(\pi, S, \text{CRS})
\]
A valid proof confirms the compliance of the update without revealing any details of \( w \). In our system, zkSNARKs are used to prove that each update to the warhead database follows the treaty’s rules-- i.e., that the transition from the prior state to the new state is valid-- without revealing the underlying state data.

\section{System Overview and Requirements}

This section introduces the design, structure, and functional goals of our warhead tracking system. The system enables secure, verifiable tracking of nuclear warheads while preserving operational confidentiality and accommodating the realities of existing infrastructure and international treaty enforcement.

We present the system in terms of its architecture, security and operational requirements, formal treaty-derived rules, setup process, data exchange protocol, and representative use cases.

\subsection{System Goals and Architecture}

The warhead tracking system is designed to allow two parties to monitor compliance with arms control treaties without disclosing sensitive military information. It achieves this through a continuous process of cryptographic commitments, zero-knowledge proofs, and selectively revealed data challenges. The confidential history of a single warhead is encapsulated in a data structure we refer to as a \textit{passport}.

Each warhead passport consists of a series of event-driven updates, where each update records operations such as movement, maintenance, or dismantlement. These updates are committed to using hash-based commitments and proven compliant with treaty rules via zkSNARKs. Crucially, commitments are unlinkable by default, hiding even the number of warheads tracked. Only through the challenge process---where selected fields and hash links are revealed---can observers begin to reconstruct partial warhead histories and the number of warheads tracked.

Each passport entry is composed of the fields listed in Table~\ref{tab:passport_fields}, which form the data structure over which commitments and zero-knowledge proofs are computed.
\begin{table}[h]
\centering
\small
\renewcommand{\arraystretch}{1.2}
\begin{tabular}{ll}
\toprule
\textbf{Field} & \textbf{Description} \\
\midrule
Date/Time & Timestamp of the event \\
Location & Geographic or facility identifier \\
Status & Warhead status (e.g., active, stored) \\
Secondary Component & Secondary warhead stage ID \\
LLC 1 & Limited-lifetime component (e.g., Tritium) \\
LLC 2 & Limited-lifetime component (e.g., Battery) \\
Operation & Operation performed (e.g., transport, repair) \\
Personnel & Responsible individual(s) \\
Exception & Boolean indicating exceptional event \\
Exception Reason & Justification for exception, if any \\
Previous Hash & Hash linking to prior update \\
\bottomrule
\end{tabular}
\caption{Full list of fields stored in each warhead passport entry.}
\label{tab:passport_fields}
\end{table}

\begin{figure}[h]
\centering
\includegraphics[width=0.9\linewidth]{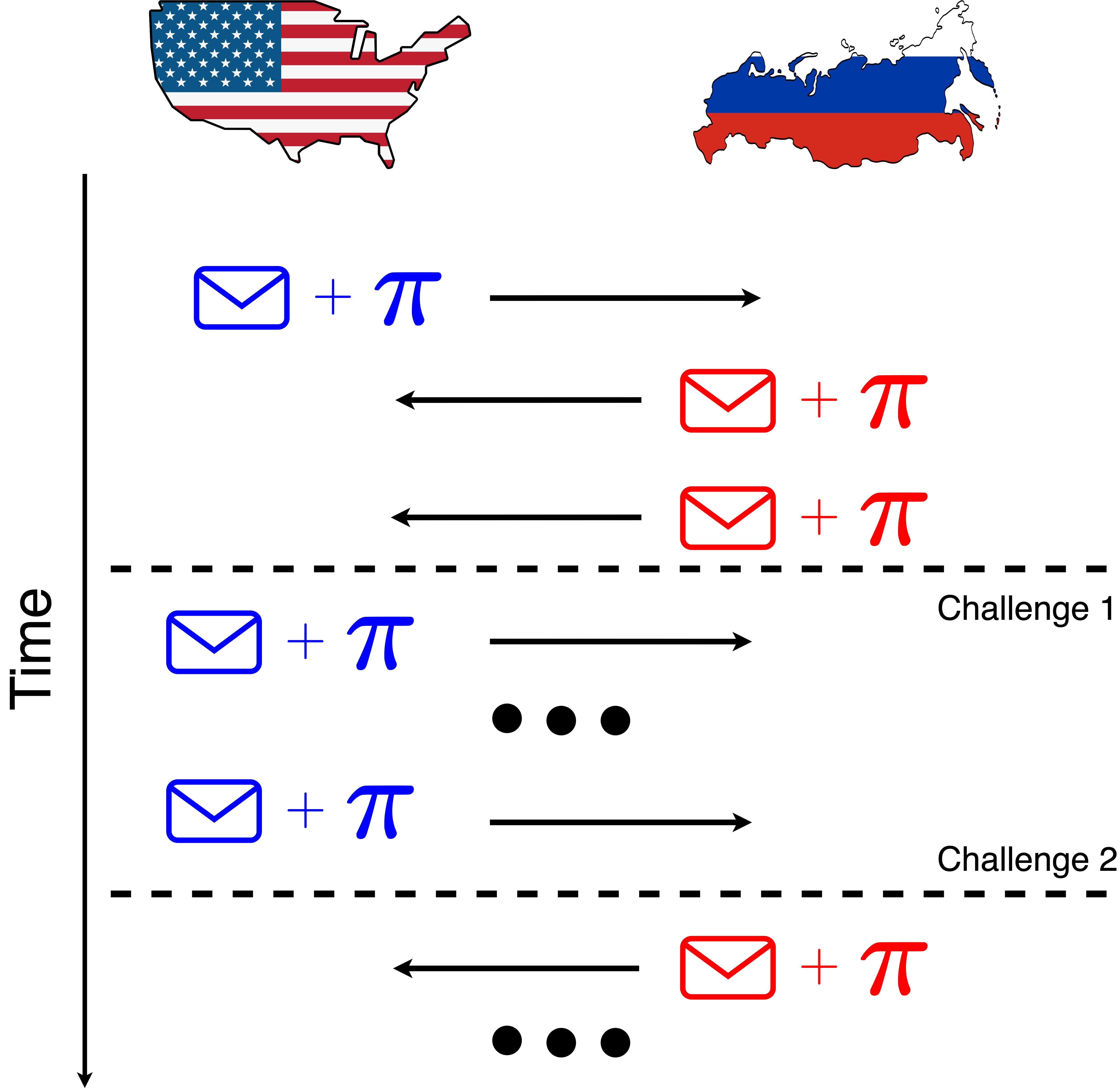}
\caption{Diagram of sharing commitments and zk proofs corresponding to updates along with data challenges over time.\protect\footnotemark}
\label{fig:system_over_time}
\end{figure}

Figure~\ref{fig:system_over_time} illustrates the temporal flow of commitments, associated zero-knowledge proofs, and challenge requests as the system operates. Initially, all commitments appear opaque; only later do specific openings expose structure.

\footnotetext{The country outlines used in the figure are from \url{https://www.vecteezy.com/free-png/usa-outline} and \url{https://www.vecteezy.com/free-png/russia}.}

\subsection{Security and Operational Requirements}

To be both technically sound and operationally viable, the system must satisfy the following requirements.

\paragraph{Security Requirements}
\begin{itemize}
\item \textbf{Confidentiality}: Warhead update data must remain hidden unless specific fields are revealed during challenge procedures.
\item \textbf{Integrity}: Commitments must be binding and unforgeable, ensuring historical updates cannot be altered.
\item \textbf{Non-Repudiation}: Parties cannot deny having submitted a commitment once it is published.
\item \textbf{Soundness}: zkSNARKs must only validate updates that conform to treaty rules.
\item \textbf{Forward Security}: Each update must link cryptographically to its predecessor, making tampering detectable.
\item \textbf{Enforceability}: Violations must be detectable and traceable to defined treaty breaches.
\end{itemize}

\paragraph{Operational Requirements}
\begin{itemize}
\item \textbf{Event-Driven Commitments}: Updates must be committed as soon as they occur.
\item \textbf{Challenge Support}: Either party must be able to challenge specific commitments at defined intervals.
\item \textbf{Efficient Verification}: Verification of zkSNARKs must be computationally lightweight.
\item \textbf{Robustness}: The system must tolerate communication failures and adversarial behavior.
\item \textbf{Integration}: The system must interoperate with existing national warhead tracking databases (DIAMONDS, AICMS) and secure communication channels \cite{gottemoeller2023analysis}.
\end{itemize}

\subsection{Treaty Rules and Event Validity}
\label{treaty_rules_and_validation}

Based on extensive discussions with recently retired personnel from the U.S. Defense Threat Reduction Agency (DTRA) directly responsible for the DIAMONDS and AICMS programs, we compiled representative treaty constraints for warhead behavior and lifecycle updates. These rules form the foundation of our zkSNARK circuits. This was key to ensure the system met the real-world constraints of the United States and Russia, was convincing to policy makers, and could be used internally by the U.S. government for testing. They include the following:
\begin{itemize}
\item Each update must include valid time, location, status, components, operations, and personnel data.
\item Time must increase monotonically and always be later than a predefined start time.
\item Transport events must have reasonable, treaty-compliant travel durations.
\item Locations, statuses, and operations must be drawn from predefined enumerated sets.
\item Transitions between operations must follow allowed sequences; some transitions are restricted.
\item Certain operations (e.g., dismantlement, storage) must occur at designated facilities.
\item Some operations must appear at least once in a valid dataset.
\item Specific operations must include changes to limited-lifetime component (LLC) values as required.
\item Transfers to road, rail, or air crews must be followed by corresponding shipment operations.
\item Exceptional events must be flagged explicitly and include a justification field.
\end{itemize}

These constraints ensure that every row in a passport reflects treaty-compliant behavior. Full rule sets for each country appear in Appendix~\ref{blue_rules} and \ref{red_rules}.

\subsection{Setup Phase}

Prior to operational use, the system undergoes a one-time setup phase:
\begin{itemize}
\item \textbf{Rule Agreement}: Treaty rules are finalized and compiled into circuits.
\item \textbf{Key Exchange}: Parties exchange cryptographic keys for commitments and verification.
\item \textbf{Public Parameters}: Agreed-upon parameters (e.g., CRS, hash functions, PRFs) are established.
\item \textbf{Initial State Commitments}: Commitments to all initial warhead records are exchanged.
\end{itemize}

This setup enables the secure exchange and verification of future updates.

\subsection{Commitment and Challenge Protocol}

After setup, the system operates in two parallel phases:

\paragraph{Data Exchange Phase} Each time a warhead undergoes an event (movement, inspection, repair, etc.), the controlling party generates a cryptographic commitment to the update and a zkSNARK proof attesting to treaty compliance. The commitment and proof are shared with the other party who immediately verifies the proof and stores the commitment.

\paragraph{Challenge Phase} At negotiated intervals, each party may select commitments from the other and request selective openings (e.g., location, operation). The responding party replies with the requested fields and a Merkle inclusion proof. If the request involves linking records, the \texttt{previous\_hash} field is revealed to establish lineage.

\begin{figure}[h]
\centering
\begin{subfigure}[b]{0.45\textwidth}
\centering
\includegraphics[width=\textwidth]{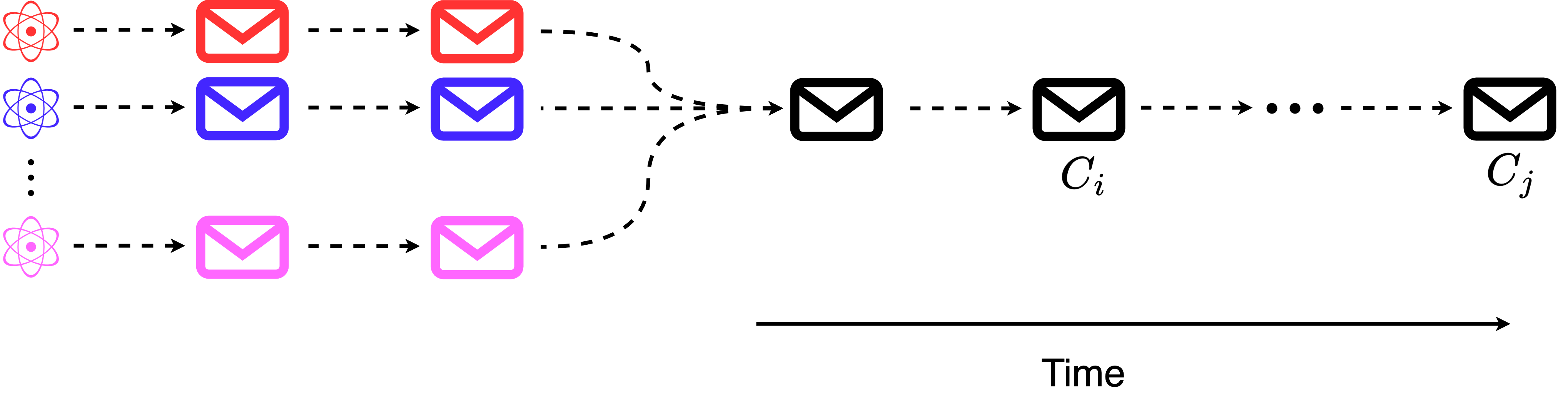}
\caption{Commitments appear opaque upon submission. Their ordering does not reveal identity.}
\label{fig:wh_example_closed}
\end{subfigure}
\hfill
\begin{subfigure}[b]{0.45\textwidth}
\centering
\includegraphics[width=\textwidth]{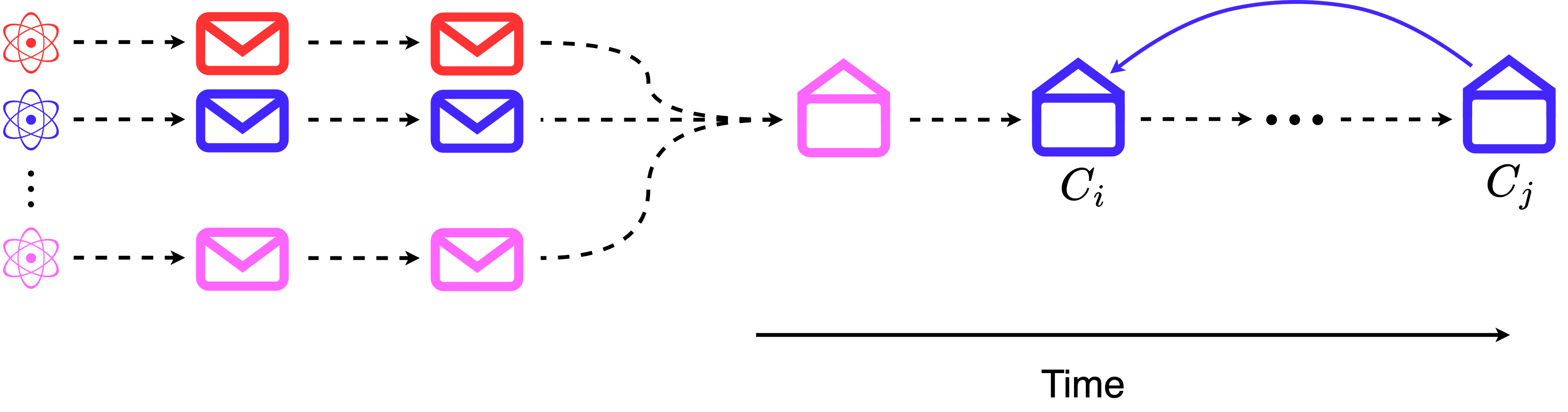}
\caption{Once previous hash fields are opened, relationships between updates are revealed. Note that this diagram has been simplified and the opener would not know that $C_i$ and $C_j$ are ``blue'' (i.e. which warhead), but simply that they are both the same color. In other words, one can learn that they pertain to the same warhead, but not the specific warhead unless the openings link them to the first commitment.}
\label{fig:wh_example_open}
\end{subfigure}
\caption{Example illustrating the relationships between open and closed commitments.}
\label{fig:wh_example}
\end{figure}

\begin{figure}[h]
\centering
\includegraphics[width=0.9\linewidth]{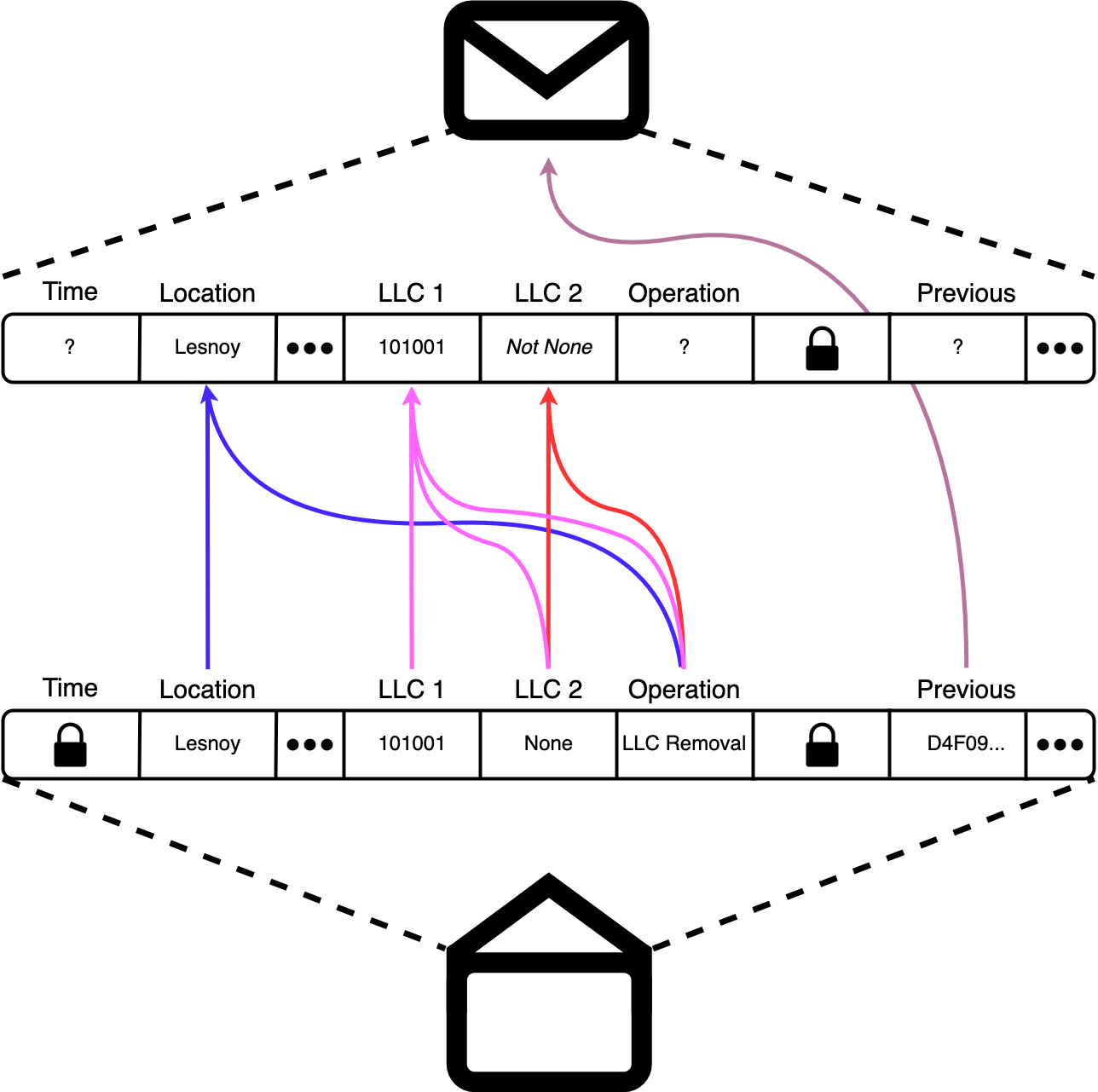}
\caption{Opening select fields of an update reveals partial information on previous updates. For example, it is possible to deduce the value of the location, LLC 1, and partial information on LLC 2 in this example. Note that the commitment is not fully opened, as fields such as Time are not revealed.}
\label{fig:commitment_discovery}
\end{figure}

Figure~\ref{fig:wh_example} illustrates how commitments are initially unlinkable (Figure~\ref{fig:wh_example_closed}) and how linkages emerge during challenges (Figure~\ref{fig:wh_example_open}). The revealed fields can give you information on the contents of the fields of another sealed update. For example, in Figure~\ref{fig:commitment_discovery}, the location, LLC 1, LLC 2, Operation, and Previous Link are revealed. Using this information, the challenging party can deduce what previous update came before this one, what its location is, what its LLC 1 value is, and that its LLC 2 value is non-empty. This is possible because the revealed operation was not a transport, so the location must remain the same. Additionally, the operation indicated that an LLC was removed. Given that the LLC 2 value is empty, the party can deduce that this is the one that was removed. Therefore, LLC 1 remains identical and LLC 2 must have been a non-empty value.

\subsection{Example Usage Scenarios}

We present two illustrative cases to demonstrate normal and exceptional system behavior.

\subsection{Example Usage}
After the completion of the setup phase where countries have agreed upon the contents of the passports, the treaty rules that passports must follow, the penalties for breaking rules, challenge time intervals/schedules, and exchanged all needed information such as identity and verification keys the following examples occur.

\subsubsection{Normal Operations with No Problems}
The United States wants to perform a maintenance operation on one of their nuclear warheads located in Pantex, Texas at time $i$. First, they mark the warhead as inactive. This triggers the conditions to share a commitment of an update with Russia. Information such as the current date and time, the location, the personnel involved, the current status (now changed to inactive), and the operation being performed are used to generate a commitment $C_i$ and zkSNARK $\pi_i$ is written to show that it follows all rules in the Treaty in relation to times prior to $i$. $C_i$ and $\pi_i$ are sent to Russia. Upon receipt, Russia verifies $\pi_i$ and stores $C_i$. Once maintenance is finished at time $j > i$, a new update is generated with corresponding information with the operation field changed to ``maintenance''. This is used to create $C_{j}$ and $\pi_{j}$ which are sent to Russia and stored and verified respectively.

Later at time $k > j$, Russia issues a data challenge for $C_{j}$, asking for the operation performed and the location. The United States responds with $\text{update}_{j}[\text{location},\allowbreak\ \text{operation},\allowbreak\ \text{previous hash}]$ and $\pi_{\text{inclusion}_{j}}$. Russia is able to verify that $\text{update}_{j}[\text{location},\allowbreak\ \text{operation},\allowbreak\ \text{previous hash}]$  was truly the location and operation performed in $C_{j}$ and therefore knows that it was a maintenance operation in Texas. Additionally, Russia now also knows that $C_j$ is the follow up to $C_i$, being in the same chain and that $C_i$ contained a status of ``inactive'' and a location of ``Pantex, TX'' due to the Treaty rules of how neighboring updates work. Note that there may have been many updates shared between time $i$ and time $j$. There is no way to learn any information on which warheads updates apply to from the commitments pertaining to updates alone. Only having the $previous\ hash$ field opened will allow you to establish that two commitments pertain to the same warhead\footnote{One exception to this is if one can somehow learn this via observations-- i.e. only one warhead has any activity going on at the time and all warheads can be observed. This extreme scenario is unrealistic, but correlations can be made based on things like satellite observations to gain confidence in suspected relationships between commitments.}.

\subsubsection{Exceptional Case Due to Natural Disaster}
Let's say that Russia plans to move a nuclear warhead from Trekhgorny to Komsomolsk-na-Amure at time $h$, which by treaty rules must take between 72-80 hours. Russia loads the warhead onto a train, which triggers the conditions to mandate sharing an update. They record the necessary information and generate commitment $C_h$ and zkProof $\pi_h$, sending them to the US. The rail transport encountered buried tracks due to a severe blizzard, delaying the shipment by 24 hours. Unfortunately, this caused the transport to take longer than the allowed time as agreed upon in the treaty rules. Upon arrival at time $k > h$, Russia marks the fields corresponding to the receipt of an inbound transport in a new update $C_k$. Critically, Russia is unable to write a proof $\pi_k$ that $C_k$ conforms to all treaty rules. Instead, they mark a special field in $C_k$ and add a description of the events with the blizzard that caused this rule violation. Russia then writes $\pi_{k}'$ that instead proves that they are asserting that an exceptional event occurred along with $update_{k}[Exception, Explanation]$ and $\pi_{inclusion_{k}}$ showing the US that it belongs to this commitment and explaining the reason to them. Upon receipt of $C_k$, $\pi_{k}'$, $update_{k}[Exception, Explanation]$, and $\pi_{inclusion_{k}}$, the US reviews the event and Exception and either accepts Russia's explanation or calls for a further examination of what took place by a bilateral consultative commission defined in the treaty.

Table~\ref{tab:warhead_passport} shows sample passport entries.

\begin{table*}[h]
\centering
\small
\renewcommand{\arraystretch}{1.2}
\begin{tabular}{llllllllll}
\toprule
\textbf{Date/Time} & \textbf{Location} & \textbf{Status} & \textbf{Secondary Comp.} & \textbf{LLC 1} & \textbf{LLC 2} & \textbf{Operation} & \textbf{Personnel} & \textbf{Previous Hash} & \textbf{Exception} \\
\midrule
11/13/2017 16:00 & CAD0L & RP & S01001 & LLC101001 & LLC201001 & R11  & AD1   &  &  \\
11/13/2017 17:00 & CAD0L & RI & S01001 & LLC101001 & LLC201001 & R21  & R63S1 & F10E... & null \\
11/14/2017 13:00 & WR63S & RI & S01001 & LLC101001 & LLC201001 & R322 & R63S1 & 2Ca0... & null \\
11/15/2017 20:30 & WR63S & RI & S01001 & LLC101001 & LLC201001 & R23  & R631  & A4D3... & null \\
\multicolumn{10}{c}{\dots\quad\dots\quad\dots} \\
\bottomrule
\end{tabular}
\caption{Example Warhead Passport Entries. Some columns are omitted for space and fields use short codes to save space as in the reference implementation.}
\label{tab:warhead_passport}
\end{table*}

\section{Construction} \label{sec:warhead-construction}

This section presents the core cryptographic construction for our warhead update tracking system. We begin with the hash function combiner used to address cross-national trust concerns, then define the commitment scheme built on Merkle trees, followed by the zkSNARK circuit enforcing treaty compliance, and conclude with the challenge protocol used to selectively audit commitments.

\subsection{Hash Functions}
Due to longstanding concerns over potential backdoors in each other's cryptographic standards, the United States and Russia do not trust a single hash function chosen by the other party. To mitigate this, we use cryptographic combiners to eliminate the need for mutual trust in a single hash function. Each side employs its own cryptographic hash function:

\begin{itemize}
\item $H_{US}$: SHA-256-- a hash function standardized by the United States.
\item $H_{RU}$: GOST R 34.11.94\footnote{This hash function can easily be swapped with other versions such as 34.11.2012. This choice was made due to availability of reference implementations to verify compatability.}-- a hash function chosen by Russia.
\end{itemize}

To ensure neither party can unilaterally compromise the integrity of the system, the combined hash function is defined as:
\[
H(x) = H_{US}(x) || H_{RU}(x)
\]

This construction guarantees that any manipulation or weakness in one function does not affect the overall security, as breaking $H(x)$ requires breaking both $H_{US}$ and $H_{RU}$.
This ensures security as long as at least one of the hash functions remains collision-resistant \cite{herzberg2005tolerant}.

\subsection{Commitment Scheme}

Each country holds a long-term secret key: $k_{US}$ for the United States and $k_{RU}$ for Russia. These keys are used with a pseudorandom function (PRF) to deterministically derive per-element randomness without storing separate round keys. Specifically, the randomness used to hide element \(j\) of the \(i\)th update is:
\[
\sigma_{t_{i,j}} = \mathsf{PRF}(k_t, (i, j)),
\]
where $t \in \{\mathsf{US}, \mathsf{RU}\}$ and $j \in \{1, \dots, n\}$. We omit the subscript $t$ when the context is clear, instead writing $\sigma_{i,j}$. Our scheme assumes the Random Oracle (RO) model and operates as follows.

\subsubsection{Update Lifecycle and Linkage}

Each update \(\mathsf{update}_i\) in a warhead's passport represents a single event in its lifecycle—such as movement, maintenance, or dismantlement. The update is a structured vector of fields, including operational metadata (e.g., timestamp, location, status), component identifiers, personnel, exception indicators, and cryptographic linkage fields. It is represented as a list of bitstrings denoted $(x_1, \dots, x_n)$ representing the $i$th update to a passport, where $n$ is a power of 2.

To ensure forward integrity and enable verifiable chains of custody, each update includes a \texttt{previous\_hash} field, which stores the Merkle root of the prior update, \(C_{i-1}\). This cryptographic chaining prevents undetected tampering with historical records: any alteration to an earlier commitment invalidates all subsequent links in the chain.

When a new event occurs, the responsible party gathers all relevant field values and populates the \(\mathsf{update}_i\) vector. Each element is then hidden with PRF-derived randomness and hashed individually (see \texttt{hide} below). These hashed elements form the leaves of a Merkle tree, whose root \(C_i\) acts as the commitment to the entire update. The country then constructs a zkSNARK proof \(\pi_i\) attesting that the transition from \(\mathsf{update}_{i-1}\) to \(\mathsf{update}_i\) conforms to treaty constraints.

\subsubsection{Commitment Function}

Each update is committed using the following three-step process:

\paragraph{Hiding Individual Elements}
Each element $x \in \mathsf{update}_i$ is committed to individually using $hide$ defined below. This allows for opening a commitment for specific elements.

\begin{lstlisting}[mathescape]
    func hide($i, j, x$):
        return $H(\sigma_{i,j} \| x)$
\end{lstlisting}

\paragraph{Constructing a Merkle Tree}
A Merkle tree is built over the resulting hidden values from $hide$. $M$ is a function that takes in a list of bitstrings and returns the resulting root of a Merkle tree. $M$ works as follows:

\begin{lstlisting}[mathescape]
    func M($(y_{1},...,y_{n})$):
        for i = 1 to n-1:
            $y_{i+n} = H(y_{2i-1},y_{2i})$
        return $y_{2n-1}$ # Merkle Tree Root
\end{lstlisting}

\paragraph{Generating a Commitment}
This is used to build the commitment scheme. $commit$ is used to commit to an update:

\begin{lstlisting}[mathescape]
    func commit(i, update):
        hidden = []
        for j = 1 to n:
            $hidden[j] = hide(i, j, update[j])$
        $C := M(hidden)$
        return $C$
\end{lstlisting}

When a country wants to share the $i$th update, it computes $C_i := \mathsf{Commit}(i, \mathsf{update}_i)$ and sends $(C_i, \pi_i)$ (defined in \ref{zkSNARK_generation}) to the other country.

\subsection{zkSNARK Generation}
\label{zkSNARK_generation}
After a country has constructed $C_i$ from $update_i$, the country constructs a zkSNARK as follows:

The witness \(w = (update_i, update_{i-1})\) consists of the current and prior updates. Treaty rules are encoded as an arithmetic circuit that validates allowed transitions between them.

A proof \(\pi \leftarrow Prove(S, w, CRS)\) is then generated using the Groth16 system \cite{groth2016size} for the treaty rules $S$, the pair of updates $w$, and the $CRS$ established during the setup phase.

When complete, the pair $(C_i, \pi_i)$ is sent to the other country.

\subsection{Challenge Phase}
To assist in confidence and trust building between the two nations, the system includes a periodic challenge protocol. Challenges occur at synchronized periodic intervals. A challenge consists of:

\begin{itemize}
\item An update index $i$.
\item A set of columns $\mathsf{cols} \subseteq \{\mathsf{col_1}, \dots, \mathsf{col_n}\}$
\end{itemize}

During the challenge phase, the country opening the commitment responds with:

\begin{itemize}
\item The values $x_j$ for all $j \in \mathsf{cols}$.
\item An inclusion proof $\pi_{inclusion}$ for all $x_j$, where $j \in \mathsf{cols}$
\end{itemize}

The verifier confirms that $\pi_{inclusion}$ indeed proves that the revealed $x_j$ were in $C_i$.
\section{Performance Analysis} \label{sec:performance}
This section evaluates the performance and feasibility of our system through a full implementation and extensive testing. We begin by describing the structure and components of our prototype, implemented in both ZoKrates and Python. We then present a detailed performance evaluation across core cryptographic operations and protocol logic, including zkSNARK proof generation and verification, Merkle tree operations, and storage requirements. Finally, we discuss the tradeoffs involved in our cryptographic hash choices and their implications for performance.

\subsection{System Implementation}

To assess feasibility, we implemented a full prototype of our protocol using both ZoKrates and Python which can be found at \url{https://github.com/NeilAPerry/Warhead-Tracking-System}. It is structured as follows:
\begin{itemize}
    \item The \textbf{ZoKrates layer} encodes the zero-knowledge circuits for treaty compliance, including all constraints needed for update validation.
    \item The \textbf{Python layer} simulates the protocol's full runtime behavior, including commitment generation, message exchange, and challenge handling.
\end{itemize}

The implementation comprises 1376 lines of Python and 2643 lines of ZoKrates, compiling to:
\begin{itemize}
    \item \textbf{23,254,511 constraints} for the Russian side, and
    \item \textbf{23,266,813 constraints} for the American side.
\end{itemize}

A significant technical contribution of our work is a complete implementation of the GOST R 34.11.94 cryptographic hash function in ZoKrates. To our knowledge, this is the first public version suitable for use in zkSNARK circuits. It enables compatibility with Russian cryptographic preferences and supports future applications requiring GOST-based commitments or hashes in zero-knowledge settings.

Beyond the zero-knowledge circuits, we implemented a Python-based manager that simulates the entire protocol workflow, including:
\begin{itemize}
\item Sending and receiving warhead update commitments.
\item Verifying zkSNARK proofs for compliance.
\item Issuing data challenges at periodic intervals.
\item Validating responses to challenges and checking inclusion proofs.
\end{itemize}
This Python framework facilitates rapid experimentation and evaluation of our protocol in a real-world setting.

\subsection{Performance Evaluation}

We evaluate the computational performance of our system across four key metrics:
\begin{itemize}
    \item Time required to generate zkSNARK proofs for treaty compliance.
    \item Time required to verify zkSNARK proofs upon receipt.
    \item Time required to generate and verify Merkle tree inclusion proofs.
    \item Storage overhead introduced by zkSNARK proofs and Merkle commitments.
\end{itemize}

These metrics reflect the protocol’s ability to support real-time commitments, rapid verification, and scalable archival of updates. It is critical that proof generation is practical on existing hardware, verification is near-instantaneous, and proof sizes remain small enough for secure transmission over existing secure channels~\cite{gottemoeller2023analysis} and are feasible to store.

Table~\ref{tab:performance_metrics} summarizes performance results across 92 test cases for the Russian side and 59 for the American side. The difference in constraint counts and test cases reflects the larger complexity of the Russian codebase. All tests were run on an AWS \texttt{m5a.4xlarge} machine with 64 GB RAM.

We observe the following:
\begin{itemize}
    \item \textbf{Witness generation} averages 4 minutes and 25 seconds per update.
    \item \textbf{Proof generation} completes in under 33 minutes.
    \item \textbf{Verification time} is under 25 milliseconds, enabling immediate compliance checks.
    \item \textbf{Proof sizes} for the zero-knowledge component are 384 bytes (uncompressed), suitable for secure transmission and indefinite storage.
\end{itemize}

\begin{table}[h]
    \centering
    \begin{tabular}{|l|c|c|}
        \hline
        Measurement (Averages) & Russian & American \\
        \hline
        Witness Calculation Time & 4 min 24 sec & 4 min 25 sec \\
        Proof Generation Time & 32 min 49 sec & 32 min 50 sec \\
        Proof Size & 384 bytes & 384 bytes \\
        Verify Time & 23.18 ms & 22.69 ms \\
        \hline
    \end{tabular}
    \caption{Performance Metrics for witness calculation, proof generation, verification time and proof sizes measured over 92 Russian and 59 American test cases.}
    \label{tab:performance_metrics}
\end{table}

In addition to the zkSNARK performance, we evaluate the runtime costs of the surrounding protocol logic implemented in Python, which manages commitments, inclusion proofs, and proof verification. These measurements reflect real-time performance of the Merkle tree–based commitment system, which runs outside the zero-knowledge layer but is essential for challenge handling and proof tracking. Table~\ref{tab:merkle_metrics} summarizes the measured performance. All metrics are averaged over 100 trials and run on an AWS \texttt{t3.xlarge} machine with 16 GB RAM.

We observe the following for the Merkle commitment layer:
\begin{itemize}
    \item \textbf{Commitment generation} averages 157 milliseconds per update.
    \item \textbf{Inclusion proof generation} completes in 591 milliseconds.
    \item \textbf{Proof verification} requires only 77 milliseconds, enabling efficient response validation.
    \item \textbf{Commitment and proof sizes} are small—79 bytes and 346 bytes respectively—supporting low-overhead transmission and storage.
\end{itemize}

\begin{table}[h]
    \centering
    \begin{tabular}{|l|c|}
        \hline
        Measurement (Average) & Value \\
        \hline
        Commitment Time & 157 ms \\
        Merkle Proof Generation Time & 591 ms \\
        Merkle Proof Verification Time & 77 ms \\
        Commitment Size & 79 bytes \\
        Proof Size & 346 bytes \\
        \hline
    \end{tabular}
    \caption{Performance of the Merkle commitments measured over 100 trials. These values reflect the average time to commit to an update, generate a Merkle proof, verify a Merkle proof, and the associated data sizes.}
    \label{tab:merkle_metrics}
\end{table}

\paragraph{Storage Requirements.}
After verification, zkSNARK proofs and Merkle inclusion proofs can be discarded. Verifiers retain only the Merkle root (79 bytes on average) and the challenge response ($\le$ 140 bytes). For a treaty involving 10,000 daily updates over 30 years, this amounts to approximately 22.3 GB of stored data in total, plus a small overhead for needed metadata. Since provers keep the original data, any discarded proofs can be regenerated if required. This design enables long-term auditability with minimal storage burden.

\paragraph{Practicality.}
These results demonstrate the system's practicality: updates can be committed and verified promptly, and communication overhead remains low. The system can thus scale to thousands of warheads and decades of historical updates without exceeding reasonable bandwidth or storage budgets.

\subsection{Hash Function Tradeoffs}
To satisfy geopolitical and cryptographic trust requirements, our system uses a dual-hash combiner: SHA-2 family (preferred by the United States) and GOST R 34.11 family (preferred by Russia). This approach improves political acceptability and offers resilience against single-hash failures, but it imposes substantial performance overhead in zero-knowledge settings. Specifically, our combined hash requires evaluating both SHA-256 and GOST per hash operation, significantly increasing the number of constraints in the zkSNARK circuit.

As a concrete comparison, a Merkle leaf proof using SHA-256 requires approximately 826,000 R1CS constraints, while a comparable Poseidon-based proof (e.g., 8:1 arity) requires only 4,050 constraints—more than 200x fewer \cite{grassi2021poseidon}. Even if we conservatively compare SHA-256 to the 2:1 Poseidon variant, the improvement is over 110x. Because our system includes both SHA-256 and GOST, the combined constraint cost is even higher. While this overhead remains tractable for treaty-scale deployment, switching to a zkSNARK-friendly hash like Poseidon~\cite{grassi2021poseidon,grassi2023poseidon2} or MiMC~\cite{albrecht2016mimc} would dramatically reduce constraint counts and proving time, with minimal changes to the surrounding protocol.
\section{Related Work}

Secure computational techniques such as zero-knowledge proofs (ZKPs) and Multiparty Computation (MPC) have been proposed to enable verifiable processes without revealing underlying data in various governmental settings such as law enforcement and the U.S. judicial system~\cite{bitan2022using,frankle2018audit}. 

In the nuclear arms control context, the aforementioned 2005 National Academy of Science report lists cryptographic techniques alongside a comprehensive list of other tools and techniques useful for monitoring compliance \cite{NAS2005}. In subsequent work, cryptographic commitments and blockchain-based escrows have been proposed as mechanisms for phased transparency in treaty declarations~\cite{philippe2019escrow, burford2020trust}.

Physical zero-knowledge verification protocols for nuclear warheads have been developed using measurements results from physical inspections~\cite{glaser2012new,glaser2014zero,kemp2016physical,engel2019physically,turturica2023homomorphic,hecla2018nuclear}. These systems prove warhead authenticity while preserving design secrecy, but often rely on specialized hardware and limited one-time measurements.

Previous studies also demonstrate the utility of cryptographic techniques for arms control in non-nuclear contexts. As far back as 1990, Richard Garwin proposed the use of cryptographic commitments for declaring and verifying NATO and Soviet conventional force deployments in Europe \cite{GarwinCFE}. Cryptographic techniques may also be a helpful tool for screening DNA synthesizers and assemblers to monitor for potential emergence of biohazards \cite{BaumDNA}, as well as for monitoring chemical trade records for compliance with the Chemical Weapons Convention \cite{Match2025}.

Finally, the cryptographic data exchange and warhead tracking system described above will necessarily complement other traditional and novel methods of nuclear arms control verification for any treaty limiting or reducing nuclear warheads. These methods and measures will involve on-site inspections and technical means of monitoring related objects and sites. As one key example of relevant work in this area, the International Partnership for Nuclear Disarmament Verification has been developing myriad tools and procedures for monitoring and inspecting nuclear weapons facilities for over a decade \cite{IPNDV2024} and testing them in notional exercises~\cite{NuDiVe2022}.

\section{Conclusion and Future Work}

We presented a cryptographic protocol for verifiable nuclear warhead tracking that balances the need for transparency with the confidentiality demands of arms control treaties. By combining cryptographic commitments, zero-knowledge proofs (zkSNARKs), and dual-trust hash combiners, our system enables treaty compliance verification without disclosing sensitive operational data. Chained commitments establish forward integrity and auditability, while periodic challenge mechanisms allow selective disclosure under controlled conditions.

Our implementation demonstrates the practical feasibility of this approach. We developed a complete prototype in ZoKrates, including a standards-conforming implementation of the GOST hash function for zkSNARK compatibility, and a Python-based simulation framework modeling real-world protocol execution. Performance evaluations show that the system can efficiently generate and verify proofs, enabling real-time operation and long-term scalability.

Future work will extend this framework to support multi-party verification, allowing any subset of $m$ out of $n$ signatories to participate in treaty enforcement. This generalization would increase flexibility in multilateral agreements and support broader coalitions. Further investigation is also needed to understand the verification infrastructure, warhead management systems, and cryptographic standards employed by other nuclear-armed states such as China and India. Adapting the protocol to interoperate with diverse operational and security frameworks will be essential for global adoption.

By uniting cryptographic rigor with real-world constraints derived from policy interviews and operational data, our protocol provides a foundation for verifiable limits and reductions of nuclear warheads. It contributes to the growing body of applications of zero-knowledge techniques in international security, where trust, secrecy, and accountability must coexist.

\paragraph{\textbf{Acknowledgments.}}
We would like to thank William Moon for helpful comments on this draft, as well as the rest of the team from the policy work done in Pomper~et al.~\cite{nonproliferationOP55Everything}. This work was funded by NSF, DARPA, State Department Verification Fund, the Simons Foundation, and the governments of Denmark, Germany, the Netherlands, Norway, and Sweden. Opinions, findings, and conclusions or recommendations expressed in this material are those of the authors and do not necessarily reflect the views of the above institutions or governments.

{\small
\bibliographystyle{abbrv}
\bibliography{reference}
}

\appendix
\section{Appendices Overview}

The appendices below contain the data classifications and corresponding rules for both sides of the U.S.-Russian cryptographic warhead tracking system. While the specific passports,  data entries, and rules are notional, the CNS project team aimed to make them realistic in accordance with actual U.S. and Russian warhead inventory management practice, to the extent that this is possible in an unclassified setting. To that end, the team referred to unclassified documentation and consulted with retired Department of Defense personnel with previous experience interacting with the U.S. and Russian warhead inventory systems. Additionally, Table~\ref{tab:warhead_passport} highlights an example selection of warhead passport entries.

The appendices cover the following information from both sides of the system:
\begin{itemize}
    \item Locations
    \item Statuses
    \item Operations
    \item Validation Rules
\end{itemize}

Possible real-world entries for each of these data fields are explained in full in the appendices and represented with notional identifiers in the passports themselves. The "Validation Rules" sections describe the relationships between consecutive passport entries that would be codified as treaty rules and used in zkSNARK circuits described in sections \ref{treaty_rules_and_validation} and \ref{zkSNARK_generation}.

Data fields that are not mentioned in the appendices do not have specific real-world counterparts and are fully notional (e.g. component and personnel identifiers).

\section{American Rules}
\label{blue_information}

This appendix contains the data classifications and corresponding rules for the American side of the system. The CNS project team compiled data fields and rules for notional U.S. warhead passports with references to the U.S. Department of Defense's 2020 \textit{Nuclear Matters Handbook} \cite{NMH2020} and in consultations with retired Defense Threat Reduction Agency (DTRA) personnel responsible for managing the DIAMONDS inventory management system.

\subsection{Locations}
Here we list the American and European locations. Many of these are partially redacted due to sensitivity \cite{NMH2020}.

\subsubsection{United States}
Here we group sites by their roles in the U.S. nuclear deterrent.

\noindent\textbf{Intercontinental Ballistic Missiles (ICBMs)}
\begin{itemize}
    \item Minot, ND
    \item Malmstrom, MT
    \item F.E. Warren, WY
\end{itemize}

\noindent\textbf{Ballistic Missile Submarines (SSBNs)}
\begin{itemize}
    \item Kitsap, WA
    \item Kings Bay, GA
\end{itemize}

\noindent\textbf{Strategic Bombers}
\begin{itemize}
    \item Minot, ND
    \item Whiteman, MO
\end{itemize}

\noindent\textbf{Logistics and Assembly}
\begin{itemize}
    \item Logistics Site: South West
    \item Assembly/Disassembly Site: Pantex, TX
\end{itemize}

\subsubsection{Europe}
Here we group sites associated with U.S. extended nuclear deterrent in Europe.

\noindent\textbf{Dual-Capable Aircraft (DCA)}
\begin{itemize}
    \item Central Europe DCA 1
    \item Central Europe DCA 2
    \item Central Europe DCA 3
    \item Southern Europe DCA 1
    \item Southern Europe DCA 2
    \item South Eastern Europe DCA 1
\end{itemize}

\subsection{Statuses}
U.S. passport status definitions are taken from the \textit{Nuclear Matters Handbook}'s Chapter 4 \cite{NMH2020}.

\begin{description}
    \item[Active Ready] ``Warheads designated available for wartime employment planning. Warheads are loaded onto missiles or available for generation on aircraft within required timelines.''
    \item[Active Hedge] ``Warheads retained for deployment to manage technological risks in the active ready stockpile or to augment the Active Ready stockpile in response to geopolitical developments. These warheads are not loaded onto missiles or aircraft. Warheads are available to deploy or upload per prescribed U.S. Strategic Command (USSTRATCOM) activation timelines.''
    \item[Active Logistics] ``Warheads used to facilitate workflow and sustain the operational status of Active Ready or Active Hedge quantities. These warheads may be in various stages of assembly in preparation for deployment. However, gas transfer systems are installed or co-located on the operational base in sufficient quantities to meet the readiness timelines specified in CCMD [Combatant Command] operational orders.''
    \item[Inactive Hedge] ``Warheads retained for deployment to manage technological risks in the Active Ready stockpile or to augment the Active Ready stockpile in response to geopolitical developments. These warheads are available to deploy or upload per prescribed USSTRATCOM activation timelines.''
    \item[Inactive Logistics] ``Warheads used for logistical and surveillance purposes. Warheads may be in various stages of disassembly.''
    \item[Inactive Reserve] ``Warheads retained to provide a long-term response for risk mitigation of technical failings in current and future LEPs [life extension programs]. Warheads in this category are exempt from future LEPs including Mods [modifications] and Alts [alterations].''
\end{description}

\subsection{Operations}
\noindent\textbf{Life Cycle Event}
The following operations correspond to key events within a warhead's life cycle, starting from a given unit's production and entry into service, and ending with retirement, disassembly into components, and dismantlement that renders the warhead inoperable.
\begin{itemize}
    \item Production  
    \item Designated for retirement  
    \item Prep for retirement  
    \item Awaiting disassembly  
    \item Awaiting dismantlement  
\end{itemize}

\noindent\textbf{Transfer of Custody}
Transfer-of-custody operations indicate when a given warhead is transferred between different authorities and personnel crews at a given site, usually during a warhead's movement. In the case of the United States, warhead movements between different bases and delivery units may involve ground, air, and/or sea segments, which are serviced by different crews.
\begin{itemize}
    \item Logistics base to transportation crew  
    \item Logistics base to air crew  
    \item Transportation crew to operations base  
    \item Operations base to delivery unit  
    \item Operations base to transportation crew  
    \item Delivery unit to operations base  
    \item Aircrew to operations base  
    \item Change of custodian  
\end{itemize}

\noindent\textbf{Transport}
Transport operations indicate the type of warhead transport (ground, air, or sea in the United States' case) and the destination of a given transport segment.
\begin{itemize}
    \item \textbf{Ground}
    \begin{itemize}
        \item To maintenance facility  
        \item To delivery system  
        \item To storage  
        \item To operations base  
        \item To transportation aircraft  
        \item To logistics base  
    \end{itemize}
    \item \textbf{Air}
    \begin{itemize}
        \item To maintenance facility  
        \item To delivery system  
        \item To storage  
        \item To operations base  
        \item To transportation aircraft  
        \item To logistics base  
    \end{itemize}
    \item \textbf{Sea}
    \begin{itemize}
        \item To maintenance facility  
        \item To delivery system  
        \item To storage  
        \item To operations base  
        \item To transportation aircraft  
        \item To logistics base  
    \end{itemize}
\end{itemize}

\noindent\textbf{Sustainment}
Sustainment operations are technical activities designed to ensure a given warhead's operational safety, security, and reliability while it remains in active service or in reserve. These operations include regular maintenance, safety, and security checks, replacement of components (including Limited Lifetime Components), and inventory surveillance to detect and correct any potential issues. 
\begin{itemize}
    \item Periodic maintenance  
    \item Maintenance check  
    \item Safety check  
    \item Security check  
    \item LLC exchange  
    \item LLC removal  
    \item Inventory  
\end{itemize}

\subsection{Validation Rules}
\label{blue_rules}

\begin{itemize}
    \item An update is valid if the new data satisfies all row validation rules.
    \item Each row consists of the following fields:
    \vspace{1em}
    \begin{description}
        \item[Time (8 bytes)] Time of Event
        \item[Location (9 bytes)] Geographic Location
        \item[Status (2 bytes)] Status of Warhead
        \item[Component (6 bytes)] Secondary warhead stage ID
        \item[LLC1 (9 bytes)] Tritium Booster Bottles
        \item[LLC2 (9 bytes)] Batteries
        \item[Operation (4 bytes)] Operation Conducted
        \item[Personnel (10 bytes)] ID of Personnel Involved
    \end{description}
    \vspace{1em}
    \item The time of the new row must be greater than the previous row.
    \item If the previous operation is a custodian change (``transfer of custody - change of custodian''), the next operation must be an inventory update (``sustainment - inventory'').
    \item If the operation is a transfer of custody, then:
    \begin{itemize}
        \item There must be two personnel entries.
        \item The two personnel entries must be distinct.
    \end{itemize}
    \item The LLC1 and LLC2 fields must not be empty unless the status is Inactive.
    \item The time field must be greater than or equal to START\_TIME. 
    \item The location must be from the predefined valid locations list.
    \item The status must be from the predefined valid statuses list.
    \item The operation must be from the predefined valid operations list.
    \item The personnel field must not be empty.
    \item The first personnel entry must not be empty.
    \item Transport time must be within the allowed range:
    \begin{itemize}
        \item If no transport event occurs, time validation is skipped.
        \item If the start and end locations are the same, any time is valid.
        \item If the operation is ``transport - to delivery system'' and the locations are the ICBM at Minot ND or Malmstrom MT, the allowed travel time is between 1 and 360 minutes.
        \item Otherwise, the allowed time interval is determined based on whether the transport is by ground, air, or sea-- where all pairs of locations have predefined allowable time intervals.
    \end{itemize}
\end{itemize}
\section{Russian Rules}
This appendix contains the data classifications and corresponding rules for the Russian side of the system. The CNS project team compiled data fields and rules for notional Russian warhead passports in consultations with retired DTRA personnel responsible for assisting their Russian counterparts in upgrading the AICMS inventory management system.

\subsection{Locations}
Here we list the Russian locations.
\subsubsection{Main Locations}
Here we group sites by listing the primary large national central site (referred to as ``S'' sites by the 12th GUMO) followed by its associated sites. Almost every associated unit corresponds to one of three main services responsible for the Russian nuclear deterrent - Strategic Rocket Forces (RF), Aerospace Forces (AF), and Navy (N). \\

\noindent\textbf{Vologda-20, Object 957 (Chebsara), units 25594, 00494}
\begin{itemize}
    \item Gatchina, Unit 44086 (AF: Tactical aviation, possibly air defense)
    \item Soltsy, Unit 75365 (AF: Long-range aviation, Tu-22M3)
    \item Kolosovka, Unit 20336 (N: Kaliningrad)
    \item Bologoye, Unit 33787 (RF: SS-25 mobile ICBMs)
    \item Teykovo, Unit 54175 (RF: SS-25, RS-24 Yars mobile ICBMs)
\end{itemize}
 
\noindent\textbf{Olenegorsk-2, Object 956 (Ramozero), unit 62834}
\begin{itemize}
    \item Gadzhiyevo, Unit 69273 (N: Northern Fleet, naval weapons, SLBMs)
    \item Severomorsk, Unit 81265 (N: Naval aviation)
    \item Zaozersk, Unit 22931 (N: Northern Fleet, naval weapons, SLBMs)
\end{itemize}

\noindent\textbf{Mozhaysk-10, Object 714, units 52025, 06031}
\begin{itemize}
    \item Tver, Unit 19089 (RF)
\end{itemize}

\noindent\textbf{Bryansk-18, Object 365 (Rzhanitsa), units 42685, 54056}
\begin{itemize}
    \item Shatalovo, Unit 23476 (AF: Tactical aviation)
    \item Kozelsk, Unit 44240 (RF: SS-19 and RS-24 Yars silo ICBMs)
    \item Shaykovka, Unit 26219 (AF: LRA, Tu-22M34)
\end{itemize}
 
\noindent\textbf{Belgorod-22, Object 1150 (Golovchino), unit 25624}
\begin{itemize}
    \item Morozovsk, Unit 55796 (AF: tactical aviation)
    \item Novorossiysk, Unit 52522 (N: Black Sea Fleet)
\end{itemize}

\noindent\textbf{Voronezh-45, Object 387 (Borisoglebsk), units 14254, 24552}
\begin{itemize}
    \item Yeysk, Unit 32161 (N: Naval aviation training centre)
\end{itemize}

\noindent\textbf{Saratov-63, Object 1050 (Krasnoarmeyskoye), units 25623, 04197}
\begin{itemize}
    \item Engels, Unit 77910 (AF: LRA, Tu-160, Tu-95MS strategic bombers)
    \item Tatischchevo, Unit 68886 (AF: SS-27 silo ICBMs)
\end{itemize}
 
\noindent\textbf{Lesnoy-4, Object 917 (Nizhnyaya Tura, formerly Sverdlovsk-45), unit 40274}
\begin{itemize}
    \item Svobodny, Unit 54203 (RF: RS-24 Yars mobile ICBMs)
\end{itemize}
 
\noindent\textbf{Trekhgorny-1, Object 936 (formerly Zlatoust-30), units 41013, 24562}
\begin{itemize}
    \item Yasny, Unit 93766 (RF: SS-18 silo ICBMs)
    \item Yushkar-Ola, Unit 54200 (RF: SS-25 mobile ICBMs)
    \item Borovsk, Anti-Ballistic Missile (ABM) Defense Site
\end{itemize}

\noindent\textbf{Irkutsk-45, Object 644 (Zalari), units 39995, 25007}
\begin{itemize}
    \item Sredniy, Unit 26221 (AF: Long-range aviation, Tu-22M3)
    \item Novosibirsk, Unit 54245 (RF: RS-24 Yars mobile ICBMs)
    \item Irkutsk, Unit 73752 (RF: SS-25 mobile ICBMs)
    \item Sibirskiy, Unit 08326 (RF: SS-25 mobile ICBMs)
    \item Solnechny, Unit 25996 (RF: SS-18 silo ICBMs)
\end{itemize}

\noindent\textbf{Komsomolsk-na-Amure-31, Object 1201 (Selikhino), units 52015, 57381}
\begin{itemize}
    \item Khurba, Unit \# unknown (AF: Tactical aviation)
    \item Ukrainka (Seryshevo), Unit 27835 (AF: Long-range aviation, Tu-95MS strategic bombers)
    \item Fokino, Unit 36199 (N: Pacific Fleet)
    \item Mongokhto, Unit 40689 (N: Naval aviation, Tu-142)
    \item Vilyuchinsk, Unit 31268 (N: Pacific Fleet, naval weapons, SLBMs)
\end{itemize}

\noindent\textbf{Khabarovsk-47, Object 1200 (Korfovskiy), units 25625, 81385}
\begin{itemize}
    \item Khabarovsk-41, Unit 23227 (Engineering)
    \item Chita-46, Unit 23233 (Engineering)
    \item Gorny, Unit 54160 (AF: Tactical aviation)
    \item Vozdvizhenka, Unit 23477 (AF: Tactical aviation)
\end{itemize}
 
\subsubsection{Rail Transfer Points}
\begin{itemize}
    \item Saratov
    \item Voronezh
    \item St. Petersburg
    \item Belgorod
    \item Komsomolsk
\end{itemize}
 
\subsubsection{Assembly/Disassembly Sites}
\begin{itemize}
    \item Trekhgorny
    \item Lesnoy
    \item Sarov (pilot site)
\end{itemize}

\subsection{Statuses}
Russian warhead status definitions are drawn from consultations with retired DTRA personnel and descriptions of the warhead life cycle in Chapter 4 of \textit{Russian Strategic Nuclear Forces} \cite{RusNuclearForces}.

\begin{description}
    \item[Production] Corresponds to both new and re-manufactured warheads that complete their assembly at the production/assembly site and are transferred to the 12th GUMO. 
    \item[Active] Indicates fully assembled warheads with all components, including limited lifetime components (LLCs), once they are transferred from production/assembly sites to storage or delivery units. Warheads remain in active service during an LLC replacement operation.
    \item[Deployed] An active warhead is considered deployed when it is installed on delivery system such as an ICBM or SLBM. Gravity bombs generally remain in storage and would not be considered deployed unless they are attached to the bomber.
    \item[Inactive] Corresponds to warheads that have Primary and Secondary components, but do not have other required components such as LLCs. Additionally, a new warhead that has not yet been transferred to the active stockpile can be considered inactive.
    \item[Reserve] Indicates warheads that have been taken out of active service but not yet expired. Reserve warheads may remain in storage, but partially disassembled, with certain components stored separately at the storage facility or at the associated Central Storage site.
    \item[Scheduled For Dismantlement] Warheads reaching the end of their life cycle are selected to be dismantled, at which point procedures to render them inoperable begin. LLCs are the first components to be removed once the designation is made, other components may be removed or disabled.
    \item[Disassembled] Indicates that a warhead no longer functions as such; almost all of its components have been separated, except perhaps the Primary and Secondary units.
    \item[Dismantled] The end of a warhead's life cycle, corresponding to final separation of Primary and Secondary components. The dismantled warhead is delivered to reprocessing for subsequent elimination or re-manufacturing, and a special commission confirms the dismantlement.
\end{description}

\subsubsection{Personnel Numbering}
Personnel numbers designate the escort from the host country responsible for overseeing a transaction. These numbers encode both the assigned site and the specific role of the escort. The numbering system works as follows:
\begin{itemize}
    \item The first 2 to 4 digits indicate the assigned site. Rail crews are assigned to rail transfer points. Road and site crews are assigned to the deployment storage site.
    \item Escorts are given numbered IDs (1, 2, 3, etc).\item Personnel conducting maintenance, inventory, and safety checks may share the same numbering format, including those involved in LLC operations. However, security check personnel are assigned a distinct numbering system.
    \item In some cases, personnel numbers may overlap across two deployment sites associated with the same central storage site. This does not present an issue, as personnel from one site are not transferred to the other.
\end{itemize}

\subsection{Operations}
\noindent\textbf{Life Cycle Event}
The following operations correspond to key events within a warhead's life cycle, starting from a given unit's production, through active service, and ending with retirement, disassembly into components, and dismantlement that renders the warhead inoperable.
\begin{itemize}
    \item Production  
    \item Designated for active service
    \item Designated for retirement  
    \item Prep for retirement  
    \item Disassembly  
    \item Dismantlement
\end{itemize}

\noindent\textbf{Transfer of Custody}
Transfer-of-custody operations indicate when a given warhead is transferred between different authorities and personnel crews at a given site, usually during a warhead's movement. In the case of Russia, many warhead movements between different storage sites and/or production facilities involve ground and rail segments, which are serviced by different crews that carry out warhead transfers at Rail Transfer Points.
\begin{itemize}
    \item Production to rail transport crew  
    \item Rail transport crew to Production  
    \item Rail transport crew to road transport crew  
    \item Road transport crew to rail transport crew  
    \item Road transport crew to storage site crew  
    \item Storage site crew to road transport crew  
    \item Road transport crew to deployment site  
    \item Deployment site to road transport crew  
    \item Production to road transport crew  
    \item Rail transport crew to storage site crew  
\end{itemize}

\noindent\textbf{Transport}
Transport operations indicate the type of warhead transport (ground or rail in Russia) and the destination of a given transport segment.
\begin{itemize}
    \item \textbf{Ground}
    \begin{itemize}
        \item To storage site  
        \item To delivery system  
        \item To rail transfer point  
        \item To Production  
    \end{itemize}
    \item \textbf{Rail}
    \begin{itemize}
        \item To Production  
        \item To rail transfer point  
    \end{itemize}
\end{itemize}

\noindent\textbf{Sustainment}
Sustainment operations are technical activities designed to ensure a given warhead's operational safety, security, and reliability while it remains in active service or in reserve. These operations include regular maintenance, safety, and security checks, replacement of components (including Limited Lifetime Components), and inventory surveillance to detect and correct any potential issues. Additionally, until Russia's suspension of the New START Treaty, warheads deployed on Russian strategic delivery systems have been subject to Re-entry Vehicle (RV) On-site Inspections - physical checks by U.S. inspectors designed to monitor compliance with the treaty.
\begin{itemize}
    \item Storage site maintenance check  
    \item Depot level maintenance at Central Storage  
    \item Safety check  
    \item Security check  
    \item LLC install or exchange  
    \item LLC removal  
    \item Inventory  
    \item Arms control RV On-site Inspection  
\end{itemize}

\subsection{Validation Rules}
\label{red_rules}
\begin{itemize}
\item An update is valid if the new data satisfies all row validation rules.
\item Each row consists of the following fields:
\vspace{1em}
\begin{description}
    \item[Time (8 bytes)] Time of Event
    \item[Location (6 bytes)] Geographic Location
    \item[Status (2 bytes)] Status of Warhead
    \item[Component (7 bytes)] Secondary warhead stage ID
    \item[LLC1 (9 bytes)] Tritium Booster Bottle ID
    \item[LLC2 (9 bytes)] Battery ID
    \item[Operation (4 bytes)] Operation Conducted
    \item[Personnel (6 bytes)] ID of Personnel Involved
\end{description}
\vspace{1em}
\item Transport time must be within the allowed range.
\item If no transport event occurs, time validation is skipped.
\item If the start and end locations are the same, any time is valid.
\item The time of the new row must be greater than or equal to the previous row.
\item LLC1 and/or LLC2 must follow expected status rules:
\begin{itemize}
\item If both LLC1 and LLC2 are empty, status must be one of: Production, Inactive, Reserve, Scheduled For Dismantlement, Disassembled, Dismantled.
\item Otherwise, the LLC status is considered valid.
\end{itemize}
\item The time field must be greater than or equal to START\_TIME. 
\item Location must be from the predefined valid locations list.
\item Status must be from the predefined valid statuses list.
\item Operation must be from the predefined valid operations list.
\item Personnel field must not be empty.
\item ``storage site maintenance check'', ``safety check'', ``security check'', and ``inventory'' operations must appear at least once in the dataset.
\item If an ``LLC install or exchange'' operation occurs, at least one of LLC1 or LLC2 must change compared to the previous row.
\item If an ``LLC removal'' operation occurs, at least one of LLC1 or LLC2 must be empty.
\item If a ``Depot level maintenance at Central Storage'' operation occurs, the location must be a central storage site (indicated by a 'C' in the second character of the location bytes).
\item The following road transfer transitions are allowed:
\begin{itemize}
\item ``storage site crew to road transport crew'' \textrightarrow{} ``(ground) to delivery system''
\item ``road transport crew to storage site crew'' \textrightarrow{} ``(ground) to storage site''
\item ``road transport crew to storage site crew'' \textrightarrow{} ``(ground) to delivery system''
\end{itemize}
\item If a previous operation was a transfer to a road crew (``rail transport crew to road transport crew'', ``storage site crew to road transport crew'', ``deployment site to road transport crew'', ``Production to road transport crew''), the next operation must be:
\begin{itemize}
\item A shipment to a rail transfer point (``(ground) to rail transfer point'', ``(rail) to rail transfer point'')
\item A shipment to a central storage site (``(ground) to storage site'')
\item A shipment to a production/disassembly site (``(ground) to Production'', ``(rail) to Production)'')
\end{itemize}
\item If the previous operation was a transfer to a rail crew (``Production to rail transport crew'' or ``road transport crew to rail transport crew''), the next operation must be a rail shipment (``(rail) to Production)'' or ``(rail) to rail transfer point'').
\end{itemize}

\end{document}